\documentclass{optica-article}

\journal{opticajournal} 

\articletype{Research Article}

\usepackage{lineno}

\begin{document}

\title{ Plasmonic Spin Meron Lattices with Height-Sensitive Topology Evolution}

\author{Anand Hegde,\authormark{1} Komal Gupta,\authormark{1}  and Chen-Bin Huang\authormark{1,*}}
\address{\authormark{1}Institute of Photonics Technologies, National Tsing Hua University, Hsinchu 30013, Taiwan}

\email{\authormark{*}robin@ee.nthu.edu.tw}

\begin{abstract*} 
We demonstrate height-controlled topological switching of plasmonic spin meron lattices above a metallic square coupling structure under circularly polarized illumination. Near the interface, an evanescent surface plasmon polariton (SPP) channel yields a N\'eel-type meron lattice with $\pm\tfrac{1}{2}$-like effective site charges. At larger heights, diffracted fields from the square edges dominate and convert the lattice into a Bloch-type configuration. Over a range of intermediate heights, crossover between the evanescent SPP and edge diffraction gives rise to rich rapid topology evolutions. The switching is accompanied by nucleation of off-boundary vortex-anti vortex pairs in the in-plane spin phase, producing height-dependent fractional site charges. Our findings are analytically formulated by linear superposition of SPPs in the plasmonic regime and Stratton-Chu model in diffraction regime and confirmed via full-wave finite-difference time-domain simulations.

\end{abstract*}

\section{Introduction}
Recent progress in structured light and nanophotonics has shown that vectorial optical fields can host spatially organized optical-spin textures. At metal interfaces, surface plasmon polaritons (SPPs) are surface-bound electromagnetic modes with evanescent decay away from the interface; they carry transverse optical spin and exhibit spin–momentum locking, enabling helicity-dependent near-field responses in the immediate vicinity of the surface \cite{bliokh2014extraordinary,aiello2015transverse,tsesses2018optical,du2019deep,dai2020plasmonic,dai2022ultrafast}. However, SPP-mediated near-field contributions decay with distance from the interface, whereas geometry-induced diffraction and radiative components persist to larger heights. Consequently, the local optical-spin texture above a finite plasmonic element is inherently height-dependent, and classifications based on a single observation plane can obscure the near-to-far evolution pathways \cite{shen2024optical}. A height-resolved description is therefore needed to separate textures dominated by near-field (evanescent) contributions from those reshaped by propagating components above finite apertures or elements.\\

In the near field above a metal interface, SPP-associated fields provide a canonical setting in which transverse optical spin arises from the coupled in-plane and surface-normal field components and their fixed phase relation~\cite{van2016universal}. This structure leads to transverse optical spin whose sign is locked to the in-plane propagation direction of the surface mode \cite{davis2020ultrafast}. In this work, we classify a meron texture as Néel-type when the in-plane optical-spin component aligns parallel or antiparallel to the local radial direction defined around an out-of-plane spin extremum, and as Bloch-type when the in-plane component circulates tangentially (perpendicular to that radial direction) \cite{nagaosa2013topological}. In SPP-associated near-field regions, the in-plane optical-spin orientation is set by the local propagation direction and the surface normal, and the resulting meron helicity is generally consistent with the Néel-type definition \cite{ghosh2021topological}.\\

Diffraction from a finite element can produce propagating field components with strong transverse gradients and a nonzero longitudinal electric-field component \cite{bliokh2015spin,bekshaev2017spin}. In the nonparaxial regime, spin–orbit interaction couples incident helicity to spatial structure, yielding structured phase and polarization distributions \cite{bliokh2011spin}. Such fields can host phase and polarization singularities that organize the local optical-spin texture in the transverse plane \cite{liu2021topological}. In particular, singularity-mediated structure can promote azimuthal circulation of the in-plane optical-spin component around out-of-plane spin extrema, consistent with the Bloch-type meron definition adopted here \cite{zhang2021bloch,zhang2022optical}. Recent free-space studies have constructed optical-spin meron lattices from the spin angular momentum vector field of diffracted beams and reported meron charges near $\pm \tfrac{1}{2}$, with the sign set by the incident helicity \cite{hegde2025geometry}.\\

Both Néel-type and Bloch-type meron textures can arise above a finite plasmonic structure because the relative influence of near-field (evanescent) and propagating field components changes with observation height \cite{cheng2025navigating}. A height-resolved analysis can therefore reveal a crossover interval in which the local optical-spin texture evolves between these two limits. Here, we investigate this evolution for a metallic square coupling structure under circularly polarized illumination, and we track the optical-spin vector field on monitor planes along z. We observe a transition from a near-field Néel-type plasmonic meron lattice to a far-field Bloch-type meron lattice, with the crossover occurring over a finite height interval that is influenced by the lateral size of the structure in our geometry. The transition is accompanied by the nucleation of opposite-vorticity vortex–antivortex pairs in the in-plane optical-spin phase, which perturbs the phase-singularity ordering of the lattice \cite{li2023simultaneous,lin2024optical}. Consistent with this defect-mediated reorganization, the site-resolved effective meron charges deviate from values near $\pm\tfrac{1}{2}$ and take height-dependent fractional values \cite{lin2021microcavity}.

\section{Conceptual framework}
\label{sec:concept_theory}

\begin{figure}[ht!]
\centering
\includegraphics[width=13cm]{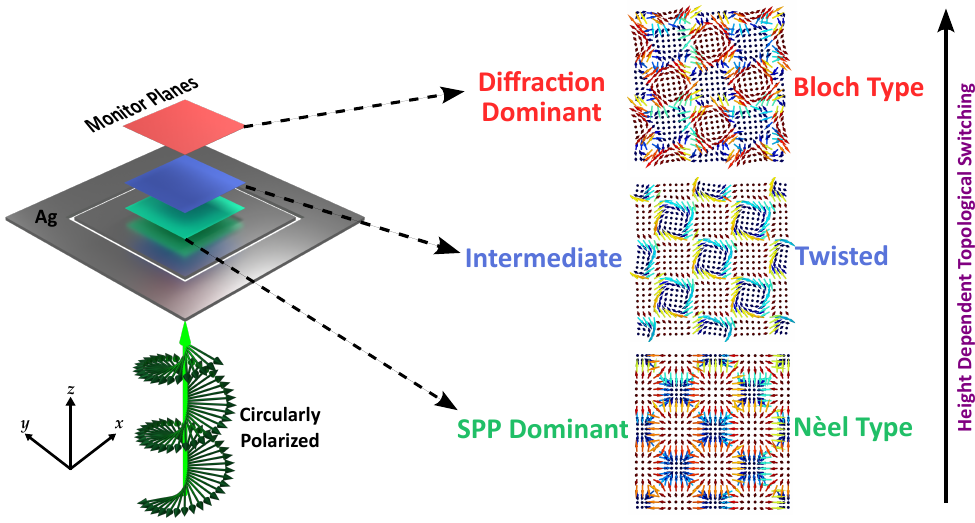}
\caption{ \textbf{Schematic of height-dependent topological switching of plasmonic spin meron lattices.} Circularly polarized illumination of a square slit in an Ag film excites SPPs and generates a diffracted field. Spin textures are evaluated on monitor planes at increasing heights above the metal-dielectric interface. In the near field, the SPP-dominant regime corresponds to a Néel-type meron lattice associated with SPP standing-wave interference and spin-momentum locking. At intermediate heights, coexistence of residual SPPs and diffracted fields yields a twisted meron lattice. In the far field, where diffraction dominates, the spin texture evolves into a Bloch-type meron lattice, indicating height-controlled topological switching.}
 \label{fig1}
\end{figure}

\begin{figure}[ht!]
\centering
\includegraphics[width=12cm]{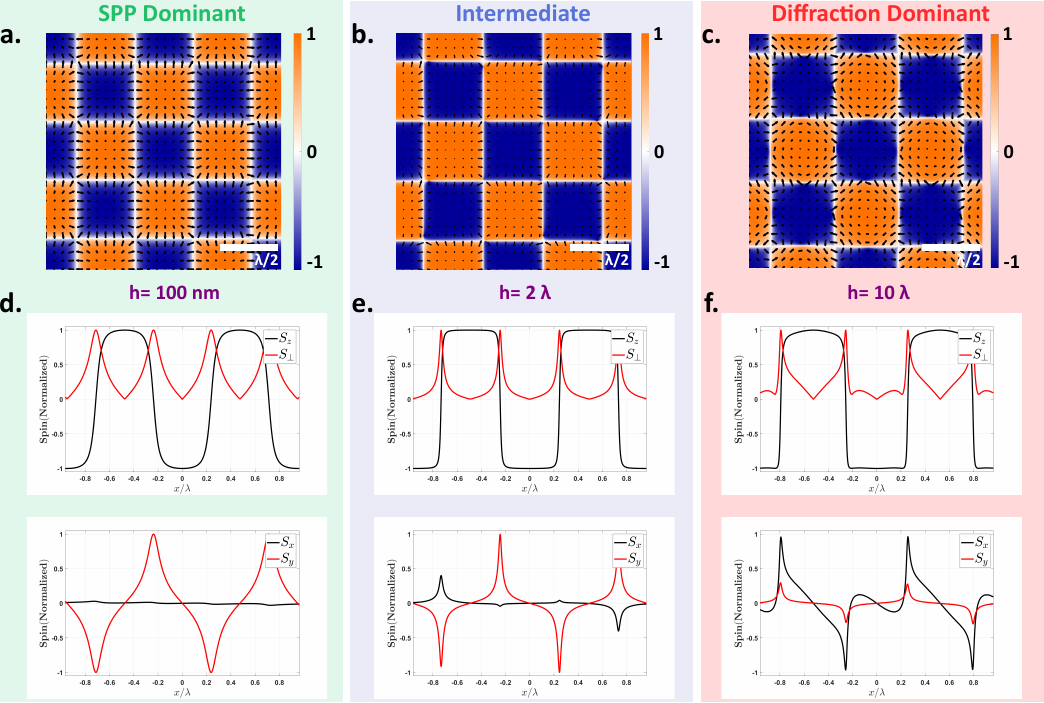}
\caption{ \textbf{Height-dependent evolution of spin structure in the $XY$ plane.} (a-c) Normalized out-of-plane spin component $S_z$ (color scale) with overlaid in-plane spin vectors $(S_x,S_y)$, evaluated at three monitor heights above the Ag surface: (a) $h=100\,\mathrm{nm}$ (SPP-dominant), (b) $h=2\lambda$ (intermediate), and (c) $h=10\lambda$ (diffraction-dominant). (d-f) Line profiles along $y=0$ quantify the redistribution of spin components with height. The upper panels compare $S_z$ and the in-plane spin magnitude $|S_{\perp}|=\sqrt{S_x^2+S_y^2}$, while the lower panels show $S_x$ and $S_y$, capturing the progression from N\'eel-like to Bloch-like in-plane rotation as the observation height increases.}
\label{fig2}
\end{figure}

We consider a square coupling structure on a metallic film illuminated by a normally incident circularly polarized plane wave, and we evaluate the scattered field in the upper half-space on monitor planes at increasing height $z$ as shown in Figure~\ref{fig1}. In the following, $z$ denotes the out-of-plane coordinate, and the discrete monitor-plane heights used in Figure~\ref{fig2} are written as $h=z$.

The total field for $z>0$ in angular-spectrum representation is,
\begin{equation}
\mathbf{E}(x,y,z)
=
\iint
\tilde{\mathbf{E}}(k_x,k_y)\,
e^{i(k_x x+k_y y)}\,
e^{i k_z z}\,
dk_x\,dk_y,
\qquad
k_z=\sqrt{k_0^2-k_x^2-k_y^2},
\label{eq:angspec_total}
\end{equation}
where $\mathbf{E}(x,y,z)$ is the complex electric field at the observation point $(x,y,z)$, $\tilde{\mathbf{E}}(k_x,k_y)$ is the angular-spectrum amplitude, $(k_x,k_y)$ are transverse wave-vector components, and $k_z$ is the longitudinal wave-vector component in the homogeneous half-space $z>0$. The free-space wave number is $k_0=2\pi/\lambda$, where $\lambda$ is the free-space wavelength. We define the transverse wave number as $k_\parallel=\sqrt{k_x^2+k_y^2}$. For $k_\parallel>k_0$, the longitudinal component is purely imaginary, $k_z=i\alpha$ with $\alpha=\sqrt{k_\parallel^2-k_0^2}>0$, and the corresponding angular-spectrum components decay as $e^{-\alpha z}$. For $k_\parallel\le k_0$, $k_z$ is real and the corresponding components propagate as $e^{ik_z z}$. This yields a spectrum-defined split of the total field into an evanescent sector and a propagating sector,
\begin{equation}
\begin{aligned}  
\mathbf{E}(x,y,z)
&=
\underbrace{
\iint_{k_\parallel>k_0}
\tilde{\mathbf{E}}(k_x,k_y)\,
e^{i(k_x x+k_y y)}\,
e^{-\alpha z}\,
dk_x\,dk_y
}_{\mathbf{E}_{\mathrm{evan}}(x,y,z)}\\
&+
\underbrace{
\iint_{k_\parallel\le k_0}
\tilde{\mathbf{E}}(k_x,k_y)\,
e^{i(k_x x+k_y y)}\,
e^{i k_z z}\,
dk_x\,dk_y
}_{\mathbf{E}_{\mathrm{prop}}(x,y,z)},
\end{aligned}
\label{eq:angspec_split_underbrace}
\end{equation}

where $\alpha=\alpha(k_\parallel)$ depends on the transverse spectral coordinate. $\mathbf{E}_{\mathrm{evan}}$ labels the evanescent-sector contribution in Eq.~(\ref{eq:angspec_split_underbrace}), which contains the SPP-near-field content above the metal interface, and $\mathbf{E}_{\mathrm{prop}}$ labels the propagating radiative contribution dominated by diffraction from the finite square. In this work, we consider $\mathbf{E}_{\mathrm{SPP}}$ for evanscent sector and $\mathbf{E}_{\mathrm{diff}}$ for the propagating sector. The height coordinate controls the balance between $\mathbf{E}_{\mathrm{SPP}}$ and $\mathbf{E}_{\mathrm{diff}}$ and produces three regimes in Figure~\ref{fig1}. \\

For the square geometry, we represent the SPP-dominant near-field content by a polygonal superposition of four edge-launched surface-wave components with polarization fixed by the SPP dispersion,
\begin{equation}
\mathbf{E}_{\mathrm{SPP}}(x,y,z)
=
e^{-\kappa z}
\sum_{n=1}^{4}
E_n(x,y)
\begin{pmatrix}
i\,\dfrac{\kappa}{|k_{\mathrm{SPP}}|}\cos\theta_n\\[6pt]
i\,\dfrac{\kappa}{|k_{\mathrm{SPP}}|}\sin\theta_n\\[6pt]
1
\end{pmatrix}
e^{i\phi_n},
\qquad
\theta_n=\frac{2\pi(n-1)}{4}.
\label{eq:Espp_polygonal_vector_square}
\end{equation}
Here $k_{\mathrm{SPP}}$ is the complex in-plane SPP wave number at the metal interface, $\kappa$ is the SPP out-of-plane decay constant in the region $z>0$, $E_n(x,y)$ is the complex in-plane envelope of the $n$th edge-launched contribution, and the phase set $\{\phi_n\}$ satisfies $\phi_n=\phi_n(\sigma)$ under the excitation convention $\sigma=\pm 1$.\\

We represent the propagating radiative contribution can be modeled using vectorial scattering theories~\cite{stratton1939diffraction}. We utilize an edge-only Stratton-Chu construction \cite{hegde2025geometry} over the four square edges to describe the total field in the radiative region,
\begin{equation}
\mathbf{E}_{\mathrm{diff}}(x,y,z)
=
\frac{1}{4\pi}
\sum_{m=1}^{4}
\int_{\Gamma_m}
\Big[
i\omega\mu_0\,G(\mathbf{r},\mathbf{r}')
(\mathbf{n}\times\mathbf{H}_{\mathrm{incident}})
+
G(\mathbf{r},\mathbf{r}')
(\mathbf{n}\times\mathbf{E}_{\mathrm{incident}})
\times\hat{\boldsymbol{\ell}}_m
\Big]
\,d\ell',
\label{eq:Ediff_SC}
\end{equation}
where Green function $G(\mathbf{r},\mathbf{r}') = \frac{e^{ik_0|\mathbf{r} \mathbf{r}'|}}{|\mathbf{r}-\mathbf{r}'|},$ with $\mathbf{r}'=(x',y',0)$, $\mathbf{r}=(x,y,z)$, $\mathbf{n}$ the surface normal, $\hat{\boldsymbol{\ell}}_m$ the unit tangent, $\Gamma_m$ the $m$th edge. The additional details about the formalism are provided in Supplementary Section 1. \\

We quantify the spin texture using the optical spin angular momentum density constructed from the complex fields,
\begin{equation}
\mathbf{s}
=
\frac{1}{4\omega}\,
\mathrm{Im}\!\left(
\varepsilon\,\mathbf{E}^\ast\times\mathbf{E}
+
\mu\,\mathbf{H}^\ast\times\mathbf{H}
\right),
\qquad
\mathbf{S}=\frac{\mathbf{s}}{|\mathbf{s}|},
\label{eq:spin_density}
\end{equation}
where $\varepsilon$ and $\mu$ denote the permittivity and permeability of the medium in the region $z>0$.The near-interface regime is SPP-dominant and inherits spin-momentum locking of the surface wave~\cite{van2016universal,dai2020plasmonic,ghosh2021topological}. The far-field regime is diffraction-dominant and inherits spin-orbit interaction of the propagating radiative field generated by the finite square~\cite{bliokh2011spin, bekshaev2017spin,hegde2025geometry}. The intermediate regime contains non-zero $\mathbf{E}_{\mathrm{SPP}}$ and $\mathbf{E}_{\mathrm{diff}}$, and the total spin texture is set by their interference. Equations~(\ref{eq:Espp_polygonal_vector_square}) and (\ref{eq:Ediff_SC}) provide a geometry-linked interpretation of Figure~\ref{fig1}: the SPP-dominant regime supports a N\'eel-type meron lattice near the interface, the diffraction-dominant regime supports a Bloch-type meron lattice at larger heights, and the meron texture switches between these limits across a finite intermediate interval where both channels contribute. \\

We validate the conceptual framework using full-wave finite-difference time-domain (FDTD) simulations. Specifically, we record the complex fields $\mathbf{E}$ and $\mathbf{H}$ on a set of monitor planes at successive heights z, evaluate the local optical spin density $\mathbf{s}$ from these complex fields, and obtain the normalized optical-spin field $\mathbf{S}$ on each plane. To promote well-ordered meron-lattice formation within a finite square aperture, we choose the side length $L$ to be in alignment with two characteristic in-plane periods: the SPP-associated interference scale set by 
$\lambda_{\mathrm{SPP}}=2\pi/\mathrm{Re}(k_{\mathrm{SPP}})$ \cite{ghosh2021topological} and the propagating diffractive scale set by the free-space wavelength $\lambda$ \cite{hegde2025geometry}. Deviations from this  square geometry can introduce boundary distortions and reduce lattice regularity. We therefore select L to satisfy an Least Common Integer-Multiple (LCIM) constraint,
\begin{equation}
L=m\lambda_{\mathrm{SPP}}=n\lambda,
\qquad m,n\in\mathbb{N},
\label{eq:LCM_constraint}
\end{equation}
which promotes coherent superposition of SPP-associated near-field interference and edge-diffracted propagating contributions across the finite domain used for height-resolved characterization. Here, simulations are performed for silver (Ag) using the optical constants in Ref. \cite{PhysRevB.6.4370} at a free-space wavelength of $\lambda=550\,\mathrm{nm}$. We choose the square side length such that $L=55\,\lambda_{\mathrm{SPP}}$ (i.e., $m=55$ and $n=53$ in Eq. \(\ref{eq:LCM_constraint}\)); unless stated otherwise, all reported numerical values in the manuscript correspond to these parameters. Further simulation parameters are provided in Supplementary Section S2, and Supplementary Section S3 discusses examples of dimensions that does not satisfy Eq. \(\ref{eq:LCM_constraint}\).\\

Figure~\ref{fig2} compares the spin components that quantify the height-driven conversion of the meron-lattice spin texture. Figures~\ref{fig2}(a-c) plot the normalized out-of-plane component $S_z$ with overlaid in-plane vectors $(S_x,S_y)$. In Figure~\ref{fig2}(a) at $h=100\,\mathrm{nm}$, $S_z$ forms a square-symmetric meron lattice and the in-plane vectors connect adjacent extrema without azimuthal circulation around the $S_z$ cores. In Figure~\ref{fig2}(b) at $h=2\lambda$, the $S_z$ extrema broaden and the in-plane vectors rotate away from the edge-to-edge alignment, indicating a change in in-plane orientation within the same lattice period. In Figure~\ref{fig2}(c) at $h=10\lambda$, $S_z$ extrema become more circularly symmetric and the in-plane vectors circulate around the $S_z$ cores, consistent with a Bloch-type rotation pattern in the transverse plane. In Figures~\ref{fig2}(d-f), the upper panels compare $S_z$ and the in-plane magnitude $|S_{\perp}|=\sqrt{S_x^2+S_y^2}$ along $y=0$, where peaks in $|S_{\perp}|$ mark the lattice-boundary regions and peaks in $S_z$ mark the lattice-core regions. In Figure~\ref{fig2}(d), the $|S_{\perp}|$ maxima occur where $S_z$ approaches zero, and the lower panel shows $S_y\approx 0$ along the cut with finite $S_x$, so the in-plane vectors on the cut are predominantly collinear, which corresponds to a N\'eel-type meron along this section. In Figure~\ref{fig2}(e), $S_z$ remains non-zero over the boundary interval where $|S_{\perp}|$ is large, and the lower panel shows finite $S_x$ and $S_y$ with unequal amplitudes and shifted extrema, which corresponds to a twisted meron configuration. In Figure~\ref{fig2}(f), the boundary peaks of $|S_{\perp}|$ overlap with non-zero $S_z$, and the lower panel shows comparable $S_x$ and $S_y$ magnitudes with alternating signs across the period, which corresponds to a Bloch-type meron configuration.

\section{Defect-mediated mechanism for height-controlled texture switching}
\label{sec:defect_mechanism}

\begin{figure}[ht!]
\centering
\includegraphics[width=13cm]{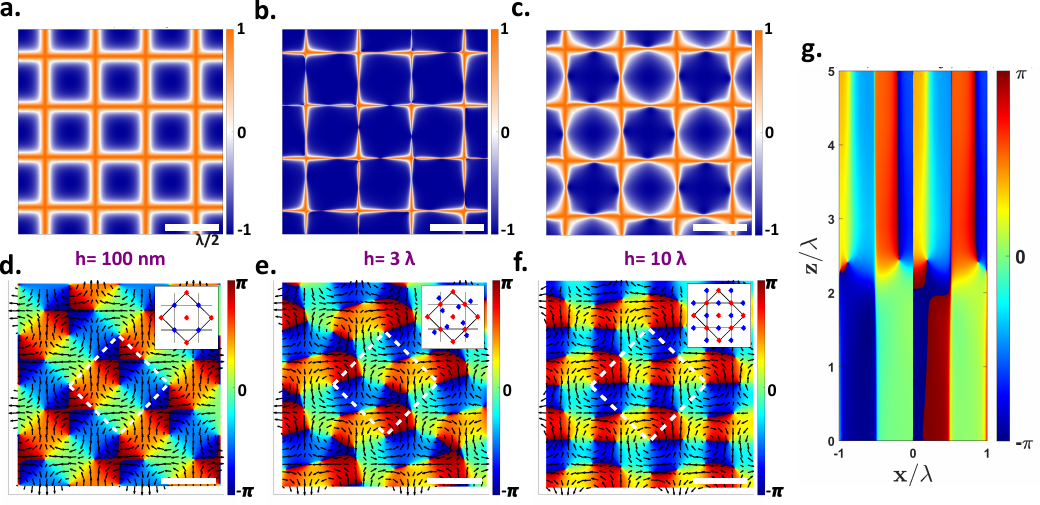}
\caption{\textbf{Defect nucleation in the in-plane spin phase across the crossover height.} (a-c) Spin-dominance indicator $\chi$ in the $XY$ plane, with $\chi\approx 0$ contours delineating the evolving lattice boundaries at (a) $h=100\,\mathrm{nm}$, (b) $h=3\lambda$, and (c) $h=10\lambda$. (d-f) In-plane spin phase $\psi=\arg(S_x+iS_y)$ at the same heights with overlaid in-plane Poynting vector, showing an ordered phase-singularity lattice in the near field (d), the emergence of off-boundary vortex-antivortex pairs in the intermediate regime (f), and the reconfigured singularity distribution in the diffraction-dominant regime (e). Insets label clockwise (red) and counterclockwise (blue) winding around representative singularities. (g) $\psi(x,z)$ map at $y=0$ showing the onset of an additional singularity within the crossover height range and its persistence for larger $z$.}
\label{fig4}
\end{figure}

The conversion of transverse spin texture with increasing height must be associated with an underlying mechanism that reorganizes the in-plane winding and drives the switching between the near-field and far-field lattice classes. We quantify the phenomenon using a spin-dominance factor $\chi$ that compares the squared magnitude of the spin component normal to a given plane with the squared magnitude of the spin components within that plane~\cite{annenkova2025universal},
\begin{equation}
\chi = |S_{\mathrm{out}}|^2 - S_{\mathrm{in}}^2,
\quad
S_{\mathrm{in}}^2 = \sum_{i\in \mathrm{plane}} S_i ^2,
\quad
|S_{\mathrm{out}}|^2 = |S_j|^2,\ \ j\perp \mathrm{plane}.
\label{eq:chi_def}
\end{equation}
Figure~\ref{fig4}(a-c) plots the $\chi$ in the $XY$ plane and uses the $\chi\approx 0$ contours to delineate the evolving lattice boundaries across height. In Figure~\ref{fig4}(a), the boundaries form a square-symmetric network that encloses well-separated $S_z$ cores. As shown in Figure~\ref{fig2}(a), at the height $h\sim2\lambda$ though the SPP fields are decayed, the topological lattice remains robust. However, in Figure~\ref{fig4}(b) , the boundary network deforms and the unit-cell edges become spatially nonuniform indicating a crossover height between $2\lambda$ and $3\lambda$. In Figure~\ref{fig4}(c), the boundary morphology differentiates between axial and diagonal sites, where axial sites approach circular contours while diagonal sites adopt higher-order polygonal outlines, which indicates that the lattice boundary is reshaped during the crossover.\\

While Figure~\ref{fig2}(a) shows that the meron cores are anchored at fixed $x$ as height increases, it is an indication that the change from N\'eel-type to Bloch-type rotation does not originate from a lateral relocation associated with $S_z$ extrema. We therefore probe the in-plane winding directly using the in-plane spin phase $\psi=\arg(S_x+iS_y)$, shown in Figure~\ref{fig4}(d-f) inspired by the previous works \cite{rybakov2025topological,wu2025photonic}. The vorticity of each phase singularity is defined by the phase winding number
\begin{equation}
v=\frac{1}{2\pi}\oint_{\mathcal{C}}\nabla\psi\cdot d\boldsymbol{\ell}
=\frac{1}{2\pi}\oint_{\mathcal{C}} d\psi,
\qquad v\in\mathbb{Z},
\label{eq:vorticity_def}
\end{equation}
where $\mathcal{C}$ is a closed loop encircling a single singularity with nonzero $|S_{\perp}|$ away from the core and $d\boldsymbol{\ell}$ is the tangential line element. In Figure~\ref{fig4}(d), the $\psi$ map contains an ordered singularity pattern within each diamond unit cell (dotted outline), and the overlaid in-plane Poynting vectors exhibit circulation with opposite handedness around distinct singularities. The inset summarizes the corresponding distribution of clockwise and counterclockwise vorticities, where the sign of $v$ is assigned by the direction of phase increase along $\mathcal{C}$. With the increasing\\

Figure~\ref{fig4}(e) shows that the ordered singularity pattern is disrupted within the crossover height range and new singularities appear away from the $\chi\approx 0$ lattice boundaries. The intermediate-height map contains paired singularities of opposite vorticity that nucleate within a single unit cell, which is the generic reaction channel of phase dislocations in nonparaxial interfering fields~\cite{karman1997creation,lin2024optical}. These off-boundary vortex-antivortex pairs redistribute the in-plane winding between the core and boundary regions and reassign the local rotation sense of $(S_x,S_y)$ around the same $S_z$ extrema. In Figure~\ref{fig4}(f), the singularity distribution reorganizes into a height-stabilized pattern in the diffraction-dominant regime \cite{lin2024optical}. The accompanied movie of $\psi$ evolution further illustrates the stability of these nucleated defects at $h=10\lambda$ through the pulse duration. The vertical evolution is resolved in Figure~\ref{fig4}(g), where $\psi(x,z)$ at $y=0$ shows the birth of an additional singularity at the previously identified crossover height and its persistence for larger $z$, which establishes defect nucleation in the in-plane spin phase as the mechanism underlying the height-controlled switching. We compare lattice site sizes and boundary deviations in the rapid conversion height zones compared to SPP and diffraction dominant zones in Supplementary S4. 

\subsection{Generalized charge accounting from polarity and vorticity}
\label{subsec:charge_accounting}

We express the site-resolved meron charge as a sum of contributions from in-plane phase singularities weighted by the local out-of-plane polarity at their cores. For a singularity indexed by $j$, we define the polarity as $p_j=\mathrm{sgn}\!\left[S_z(\mathbf{r}_j)\right]~\mathrm{if}~\in\{-1,+1\}~\mathrm{|S_z|\neq 0}~\mathrm{else}~p_j=0$ where the singularity position $\mathbf{r}_j$, and we use the vorticity $v_j\in\mathbb{Z}$ from Eq.~(\ref{eq:vorticity_def}); the generalized per-site charge is then
\begin{equation}
Q_{\mathrm{site}}
=
Q_{\mathrm{CW}}+Q_{\mathrm{CCW}}
=
\frac{1}{2}\sum_{j\in \mathrm{site}} p_j v_j,
\label{eq:Q_general}
\end{equation}

where the partition into $Q_{\mathrm{CW}}$ and $Q_{\mathrm{CCW}}$ groups terms by the circulation sense identified in Figure~\ref{fig4}(d-f). In Figure~\ref{fig4}(d), each lattice site contains a single core-associated singularity with $p_j\neq 0$, while the boundary singularities satisfy $p_j=0$ because $S_z$ vanishes on the $\chi\approx 0$ contours, so Eq.~(\ref{eq:Q_general}) reduces to a half-integer site charge set by the core polarity and vorticity. In Figure~\ref{fig4}(e), the diffracted-field contribution perturbs the singularity ordering and drives a splitting of counterclockwise singularities into vortex-antivortex pairs with opposite $v_j$, followed by migration within the unit cell, so additional terms with $p_j\neq 0$ enter the sum in Eq.~(\ref{eq:Q_general}). This defect-mediated redistribution yields fractional $Q_{\mathrm{site}}$ values whose magnitude varies with height, and the height dependence is quantified in the topology analysis where the site charges are extracted from the full skyrmion-density maps. The deviation from half integer values are also reported in literature in certain non-photonic systems (for example: \cite{mukai2022skyrmion,bazeia2021configurational}).

\subsection{A note on controllability of the critical height and sample size}
The critical height region is dependent on the coupling structure size.  However, it is worth noting that the topological structure associated with SPPs persists even after SPP fields have diminished to near-zero values unless the some perturbation forces the change. At every given height, the strength of the perturbation due to the encroaching diffracted field at the monitor plane depends on the lateral size of the sample. For the coupling structure studied the critical height of conversion is at $2\lambda<h_{\mathrm{critical}}<3\lambda$. Below the critical height there is no rotation characteristics of spin found at $y=0$ plane. After critical height the rotation of in-plane spin components become prominent exhibiting Bloch type behavior. The detailed analysis on the determination of critical height can be found in Supplementary S5. \\

 When the coupling structure size is larger compared to the one studied here the critical height region appears at a larger height while structure lateral size smaller it appears at the smaller heights. For lateral sizes smaller than the sizes studied in this manuscript ($\tfrac{L}{\lambda}<<55$), the diffracted field strength is stronger at a given monitor height. This creates well formed Bloch merons at a much shorter critical height ($\ll2\lambda$). We have observed for a structure of $8\lambda_{\mathrm{SPP}}\times8\lambda_{\mathrm{SPP}}$ size, the critical height range is between $400~\mathrm{nm}$ and $500~\mathrm{nm}$. The details can be found in Supplementary section S6.

\begin{figure}[ht!]
\centering
\includegraphics[width=6cm]{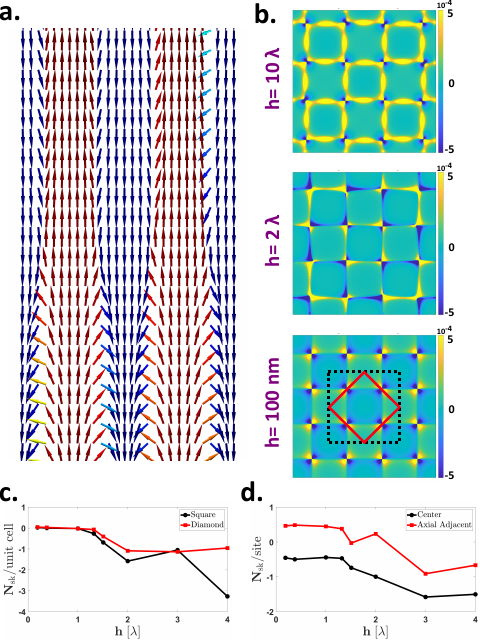}
\caption{\textbf{Height-controlled topological switching of the meron lattice.} (a,b) Skyrmion-density maps in the $XY$ plane at representative heights in the SPP-dominant and diffraction-dominant regimes. (c) Total skyrmion number $N_{\mathrm{Sk}}$ versus height $h$ evaluated for a square unit cell containing $2\times 2$ sites and for a diamond unit cell containing two sites. (d) Site-resolved $N_{\mathrm{Sk}}$ versus height for a central site and its axial-adjacent site, which carry equal-magnitude and opposite-sign values in the SPP-dominant regime and undergo switching across the crossover height range.}
\label{fig5}
\end{figure}
\section{Skyrmion Number Evaluation}
While in-plane singularity countings at each height are not easily accessible, we quantify the height-dependent topology of the spin texture using the skyrmion number $N_{\mathrm{Sk}}=\iint_{\mathcal{A}} D(x,y)\,dx\,dy$ evaluated over a chosen unit-cell area $\mathcal{A}$ in the $XY$ plane, where the skyrmion density is defined as
\begin{equation}
D(x,y)
=
\frac{1}{4\pi}\,
\mathbf{S}(x,y)\cdot
\left(
\frac{\partial \mathbf{S}(x,y)}{\partial x}
\times
\frac{\partial \mathbf{S}(x,y)}{\partial y}
\right).
\label{eq:sk_density}
\end{equation}
Figure~\ref{fig5} summarizes the height-controlled evolution of $D(x,y)$ and the resulting $N_{\mathrm{Sk}}$ extracted from the FDTD spin field $\mathbf{S}(x,y)$. Figure~\ref{fig5}(a) shows a lateral view of the three-dimensional spin texture that visualizes the near-field N\'eel-type configuration through predominantly radial in-plane orientation around the $S_z$ cores and its progressive reorientation above the crossover height. Figures~\ref{fig5}(b) compares the skyrmion-density distributions at representative heights and shows that the positive and negative density lobes redistribute across the unit cell as height increases. Figure~\ref{fig5}(c) plots the integrated $N_{\mathrm{Sk}}$ versus height for a square unit cell containing $2\times 2$ sites and for a diamond unit cell containing two sites as shown in the bottom panel of the Figure~\ref{fig5}(b). In the SPP-dominant regime, the site-resolved charges within each unit cell have equal magnitude and opposite sign, so the net $N_{\mathrm{Sk}}$ of the unit cell is near zero. Within the crossover height range, $N_{\mathrm{Sk}}$ deviates from zero, which indicates that the per-site charges no longer follow half-integer cancellation within the unit cell. Figure~\ref{fig5}(d) resolves this deviation by plotting $N_{\mathrm{Sk}}$ per site for a central site and its axial-adjacent site. In the SPP-dominant regime these two sites carry opposite sign, and across the crossover interval the axial-adjacent site switches sign and approaches the same sign as the central site, while the central-site magnitude departs from the half-integer value.

\section{Conclusion}
We demonstrated height-controlled switching of plasmonic spin meron lattices above a finite metallic square under circularly polarized excitation. A spectrum-defined decomposition of the scattered field establishes two competing channels with distinct height scaling: an evanescent sector with SPP contribution that dominates near the interface and a propagating field contribution involving diffraction that dominates at larger heights. FDTD spin maps resolve three regimes, with a N\'eel-type meron lattice in the SPP-dominant near field, a twisted lattice within a narrow crossover interval, and a Bloch-type meron lattice in the diffraction-dominant region. The transition occurs rapidly with changing height demanding a critical height region which is coupling structure size dependent. The conversion is mediated by defect nucleation in the in-plane spin phase, where off-boundary vortex-antivortex pairs redistribute winding and drive deviations of site-resolved charges from half-integer values to fractional, height-dependent values. The crossover height and the charge plateaus depend on the finite lateral size through the relative spectral weights of the two channels, providing a geometry-based route to design and control topological spin textures in plasmonic near-to-far fields.


\begin{backmatter}
\bmsection{Funding}
This work was supported by the National Science and Technology Council, Taiwan under Grants 113-2221-E-007-066-MY2 and 114-2112-M-007-007-.

\bmsection{Acknowledgment}
The authors thank Han-Ting Lin for laying the preliminary groundwork for this conceptualization and insightful discussions.

\bmsection{Disclosures}
The authors declare no conflicts of interest.

\bmsection{Data Availability Statement}
Data underlying the results presented in this paper are not
publicly available at this time but may be obtained from the authors upon reason-
able request.

\bmsection{Supplemental document}
See Supplemental Information for supporting content.

\end{backmatter}

\bibliography{sample}

@PREAMBLE{
 "\providecommand{\noopsort}[1]{}" 
 # "\providecommand{\singleletter}[1]{#1}%" 
}

@article{shen2024optical,
  title={Optical skyrmions and other topological quasiparticles of light},
  author={Shen, Yijie and Zhang, Qiang and Shi, Peng and Du, Luping and Yuan, Xiaocong and Zayats, Anatoly V},
  journal={Nature Photonics},
  volume={18},
  number={1},
  pages={15--25},
  year={2024},
  publisher={Nature Publishing Group UK London}
}

@article{tsesses2018optical,
  title={Optical skyrmion lattice in evanescent electromagnetic fields},
  author={Tsesses, S and Ostrovsky, E and Cohen, K and Gjonaj, B and Lindner, NH and Bartal, G},
  journal={Science},
  volume={361},
  number={6406},
  pages={993--996},
  year={2018},
  publisher={American Association for the Advancement of Science}
}

@article{du2019deep,
  title={Deep-subwavelength features of photonic skyrmions in a confined electromagnetic field with orbital angular momentum},
  author={Du, Luping and Yang, Aiping and Zayats, Anatoly V and Yuan, Xiaocong},
  journal={Nature Physics},
  volume={15},
  number={7},
  pages={650--654},
  year={2019},
  publisher={Nature Publishing Group UK London}
}

@article{dai2020plasmonic,
  title={Plasmonic topological quasiparticle on the nanometre and femtosecond scales},
  author={Dai, Yanan and Zhou, Zhikang and Ghosh, Atreyie and Mong, Roger SK and Kubo, Atsushi and Huang, Chen-Bin and Petek, Hrvoje},
  journal={Nature},
  volume={588},
  number={7839},
  pages={616--619},
  year={2020},
  publisher={Nature Publishing Group UK London}
}

@article{nagaosa2013topological,
  title={Topological properties and dynamics of magnetic skyrmions},
  author={Nagaosa, Naoto and Tokura, Yoshinori},
  journal={Nature nanotechnology},
  volume={8},
  number={12},
  pages={899--911},
  year={2013},
  publisher={Nature Publishing Group UK London}
}

@article{zhang2022optical,
  title={Optical topological lattices of Bloch-type skyrmion and meron topologies},
  author={Zhang, Qiang and Xie, Zhenwei and Shi, Peng and Yang, Hui and He, Hairong and Du, Luping and Yuan, Xiaocong},
  journal={Photonics Research},
  volume={10},
  number={4},
  pages={947--957},
  year={2022},
  publisher={Chinese Laser Press and Optica Publishing Group}
}

@article{ghosh2021topological,
  title={A topological lattice of plasmonic merons},
  author={Ghosh, Atreyie and Yang, Sena and Dai, Yanan and Zhou, Zhikang and Wang, Tianyi and Huang, Chen-Bin and Petek, Hrvoje},
  journal={Applied Physics Reviews},
  volume={8},
  number={4},
  year={2021},
  publisher={AIP Publishing}
}

@article{dai2022ultrafast,
  title={Ultrafast microscopy of a twisted plasmonic spin skyrmion},
  author={Dai, Yanan and Zhou, Zhikang and Ghosh, Atreyie and Kapoor, Karan and D{\k{a}}browski, Maciej and Kubo, Atsushi and Huang, Chen-Bin and Petek, Hrvoje},
  journal={Applied Physics Reviews},
  volume={9},
  number={1},
  year={2022},
  publisher={AIP Publishing}
}

@article{bliokh2015spin,
  title={Spin--orbit interactions of light},
  author={Bliokh, Konstantin Yu and Rodr{\'\i}guez-Fortu{\~n}o, Francisco J and Nori, Franco and Zayats, Anatoly V},
  journal={Nature Photonics},
  volume={9},
  number={12},
  pages={796--808},
  year={2015},
  publisher={Nature Publishing Group}
}

@article{stratton1939diffraction,
  title={Diffraction theory of electromagnetic waves},
  author={Stratton, Julius Adams and Chu, LJ},
  journal={Physical review},
  volume={56},
  number={1},
  pages={99},
  year={1939},
  publisher={APS}
}

@article{davis2020ultrafast,
  title={Ultrafast vector imaging of plasmonic skyrmion dynamics with deep subwavelength resolution},
  author={Davis, Timothy J and Janoschka, David and Dreher, Pascal and Frank, Bettina and Meyer zu Heringdorf, Frank-J and Giessen, Harald},
  journal={Science},
  volume={368},
  number={6489},
  pages={eaba6415},
  year={2020},
  publisher={American Association for the Advancement of Science}
}

@article{bliokh2014extraordinary,
  title={Extraordinary momentum and spin in evanescent waves},
  author={Bliokh, Konstantin Y and Bekshaev, Aleksandr Y and Nori, Franco},
  journal={Nature communications},
  volume={5},
  number={1},
  pages={3300},
  year={2014},
  publisher={Nature Publishing Group UK London}
}

@article{lin2021microcavity,
  title={Microcavity-based generation of full Poincar{\'e} beams with arbitrary skyrmion numbers},
  author={Lin, Wenbo and Ota, Yasutomo and Arakawa, Yasuhiko and Iwamoto, Satoshi},
  journal={Physical Review Research},
  volume={3},
  number={2},
  pages={023055},
  year={2021},
  publisher={APS}
}

@article{aiello2015transverse,
  title={From transverse angular momentum to photonic wheels},
  author={Aiello, Andrea and Banzer, Peter and Neugebauer, Martin and Leuchs, Gerd},
  journal={Nature Photonics},
  volume={9},
  number={12},
  pages={789--795},
  year={2015},
  publisher={Nature Publishing Group UK London}
}

@article{van2016universal,
  title={Universal spin-momentum locking of evanescent waves},
  author={Van Mechelen, Todd and Jacob, Zubin},
  journal={Optica},
  volume={3},
  number={2},
  pages={118--126},
  year={2016},
  publisher={Optical Society of America}
}

@article{zhang2021bloch,
  title={Bloch-type photonic skyrmions in optical chiral multilayers},
  author={Zhang, Qiang and Xie, Zhenwei and Du, Luping and Shi, Peng and Yuan, Xiaocong},
  journal={Physical Review Research},
  volume={3},
  number={2},
  pages={023109},
  year={2021},
  publisher={APS}
}

@article{bekshaev2017spin,
  title={Spin--orbit interaction of light and diffraction of polarized beams},
  author={Bekshaev, Aleksandr Ya},
  journal={Journal of Optics},
  volume={19},
  number={8},
  pages={085602},
  year={2017},
  publisher={IOP Publishing}
}

@article{liu2021topological,
  title={Topological polarization singularities in metaphotonics},
  author={Liu, Wenzhe and Liu, Wei and Shi, Lei and Kivshar, Yuri},
  journal={Nanophotonics},
  volume={10},
  number={5},
  pages={1469--1486},
  year={2021},
  publisher={De Gruyter}
}

@article{hegde2025geometry,
  title={Geometry-driven lattice of photonic spin-meron tubes in free space},
  author={Hegde, Anand and Gupta, Komal and Dai, Yanan and Huang, Chen-Bin},
  journal={arXiv preprint arXiv:2508.14450},
  year={2025}
}

@article{bliokh2011spin,
  title={Spin-to-orbital angular momentum conversion in focusing, scattering, and imaging systems},
  author={Bliokh, Konstantin Y and Ostrovskaya, Elena A and Alonso, Miguel A and Rodr{\'\i}guez-Herrera, Oscar G and Lara, David and Dainty, Chris},
  journal={Optics express},
  volume={19},
  number={27},
  pages={26132--26149},
  year={2011},
  publisher={OSA}
}

@article{cheng2025navigating,
  title={Navigating optical skyrmions—from historical origins to applications: tutorial},
  author={Cheng, Cheng and Rao, Lixi and Ye, Junyi and Zhao, <? Tex$\backslash$break?> Xingqi and Che, Zhiyuan and Liu, Wenzhe and Wang, <? Tex$\backslash$break?> Jiajun and Shi, Lei},
  journal={Advances in Optics and Photonics},
  volume={18},
  number={1},
  pages={1--105},
  year={2025},
  publisher={Optica Publishing Group}
}

@article{li2023simultaneous,
  title={Simultaneous creation of multiple vortex-antivortex pairs in momentum space in photonic lattices},
  author={Li, Feng and Koniakhin, Sergei V and Nalitov, Anton V and Cherotchenko, Evgeniia and Solnyshkov, Dmitry D and Malpuech, Guillaume and Xiao, Min and Zhang, Yanpeng and Zhang, Zhaoyang},
  journal={Advanced Photonics},
  volume={5},
  number={6},
  pages={066007--066007},
  year={2023},
  publisher={Society of Photo-Optical Instrumentation Engineers}
}

@article{lin2024optical,
  title={Optical vortex-antivortex crystallization in free space},
  author={Lin, Haolin and Liao, Yixuan and Liu, Guohua and Ren, Jianbin and Li, Zhen and Chen, Zhenqiang and Malomed, Boris A and Fu, Shenhe},
  journal={Nature Communications},
  volume={15},
  number={1},
  pages={6178},
  year={2024},
  publisher={Nature Publishing Group UK London}
}

@article{annenkova2025universal,
  title={Universal nondiffractive topological spin textures in vortex cores of light and sound},
  author={Annenkova, Elena and Afanasev, Andrei and Brasselet, Etienne},
  journal={arXiv preprint arXiv:2512.02964},
  year={2025}
}

@article{wu2025photonic,
  title={Photonic torons with 3D topology transitions and tunable spin monopoles},
  author={Wu, Haijun and Mata-Cervera, Nilo and Wang, Haiwen and Zhu, Zhihan and Qiu, Chengwei and Shen, Yijie},
  journal={Physical Review Letters},
  volume={135},
  number={6},
  pages={063802},
  year={2025},
  publisher={APS}
}

@article{rybakov2025topological,
  title={Topological invariants of vortices, merons, skyrmions, and their combinations in continuous and discrete systems},
  author={Rybakov, Filipp N and Eriksson, Olle and Kiselev, Nikolai S},
  journal={Physical Review B},
  volume={111},
  number={13},
  pages={134417},
  year={2025},
  publisher={APS}
}

@article{mukai2022skyrmion,
  title={Skyrmion and meron ordering in quasi-two-dimensional chiral magnets},
  author={Mukai, Natsuki and Leonov, Andrey O},
  journal={Physical Review B},
  volume={106},
  number={22},
  pages={224428},
  year={2022},
  publisher={APS}
}

@article{bazeia2021configurational,
  title={Configurational entropy of skyrmions and half-skyrmions in planar magnetic elements},
  author={Bazeia, D and Rodrigues, EIB},
  journal={Physics Letters A},
  volume={392},
  pages={127170},
  year={2021},
  publisher={Elsevier}
}

@article{karman1997creation,
  title={Creation and annihilation of phase singularities in a focal field},
  author={Karman, GP and Beijersbergen, MW and Van Duijl, A and Woerdman, JP},
  journal={Optics letters},
  volume={22},
  number={19},
  pages={1503--1505},
  year={1997},
  publisher={Optical Society of America}
}

@article{PhysRevB.6.4370,
  title = {Optical Constants of the Noble Metals},
  author = {Johnson, P. B. and Christy, R. W.},
  journal = {Phys. Rev. B},
  volume = {6},
  issue = {12},
  pages = {4370--4379},
  numpages = {0},
  year = {1972},
  month = {Dec},
  publisher = {American Physical Society},
  doi = {10.1103/PhysRevB.6.4370},
  url = {https://link.aps.org/doi/10.1103/PhysRevB.6.4370}
}

\end{document}